\documentclass[12pt,letterpaper]{JHEP3}

\usepackage{graphicx}
\usepackage{amsmath}

\def\cN{\mathcal{N}}
\def\cL{\mathcal{L}}
\def\bR{\mathbb{R}}
\def\bx{\mathbf{x}}

\def\Lie{\pounds}
\def\lquote{{\vphantom{\big|}}^{\hbox{\textbf{``}}}}
\def\rquote{\,{\vphantom{\big|}}^{\hbox{\textbf{''}}}}
\def\sumq{s}
\let\citen\cite
\immediate\write18{../../scripts/bibgen \jobname.tex}

\preprint{
\hbox{}\hfill TIT/HEP-570\\ 
\hbox{}\hfill arXiv:0704.1819}

\title{
Comments on  Charges and Near-Horizon Data of Black Rings
}

\author{
 Kentaro {Hanaki}$^1$,
Keisuke {Ohashi}$^2$
and Yuji {Tachikawa}$^3$\\

\bigskip

$^1$ Department of Physics, University of Michigan,\\
	Ann Arbor, MI 48109-1120, USA\\
	E-mail: \email{hanaki@umich.edu}
\medskip

$^2$ DAMTP, Centre for Mathematical Sciences, Cambridge University, \\
	Wilberforce Road, Cambridge CB3OWA, UK \\
	E-mail: \email{keisuke@th.phys.titech.ac.jp}
\medskip

$^3$ School of Natural Sciences, Institute for Advanced Study,\\
 Princeton,  New Jersey 08540, USA\\
 	E-mail: \email{yujitach@ias.edu}
}

\abstract{
We study how the charges of the black rings
measured at the asymptotic infinity
are encoded in the near-horizon metric and gauge potentials,
independent of the detailed structure of the connecting region.
Our analysis clarifies how
different sets of four-dimensional charges
can be assigned to a single five-dimensional object 
under the Kaluza-Klein reduction.
Possible choices
are related by the Witten effect on dyons and by a large gauge transformation
in four and five dimensions, respectively.
}


\begin{document}
\section{Introduction}
\label{sec:introduction}

One of the achievements of  string/M theory is 
the microscopic explanation for the Bekenstein-Hawking
entropy for a class of four-dimensional supersymmetric
black holes \cite{SV,MSW}.
The microscopic counting  predicts subleading corrections to the entropy,
which can also be calculated from the macroscopic point of view,
i.e.~from stringy modifications to 
the Einstein-Hilbert Lagrangian \cite{MoReview}. 
Comparison of the two approaches 
has proven to be very fruitful,
e.g.~it has led to the relation to the partition function of topological strings \cite{OSV}.
Beginning in Ref.~\citen{entropySen},
it has been  also generalized  to non-supersymmetric extremal black holes
using the fact that the near-horizon geometry has enhanced symmetry.
The analysis has also been extended to rotating black holes \cite{rotatingattractor}.

There is a richer set of supersymmetric black objects in five dimensions, 
including black rings \cite{ring}, on which we focus.
The entropy is still given by the area law macroscopically to leading order,
and it can be understood microscopically using a D-brane construction 
\cite{microring1,microring2}.   
The understanding of higher-derivative corrections
remains more elusive \cite{microhigher1,microhigher2,CaiPang}.
One reason for this is that the supersymmetric higher-derivative
terms were not known until quite recently \cite{HOT}.
Even with this supersymmetric higher-derivative action,
it has been quite difficult to construct the black ring solution
embedded in the asymptotically flat spacetime,
and it is preferable if we can only  study the near horizon geometry.
Then the problem is to find the charges carried by the black ring
from its data at the near-horizon region.

The usual approach taken in the literature so far is to consider the
dimensional reduction along a circle down to four dimensions,
and to study the charges there \cite{CaiPang,Morales,Cardoso,ringrule}. 
Then, the attractor mechanism fixes
the scalar vacuum expectation values (vevs) and the metric at the horizon
by the electric and magnetic charges \cite{4dattractor1,4dattractor2}.
Conversely, the magnetic charge can be measured by the flux,
and the electric charge can be found by taking the variation
of the Lagrangian by the gauge potential. 
In this way, the entropy as a function of charges
can be obtained from the analysis of  
the near-horizon region alone \cite{entropySen,rotatingattractor}.
Nevertheless, it has not been clarified how to reconcile the
competing proposals \cite{microring1,microring2,taub1,taub2,taub3}
of the mapping between the four- and five-dimensional charges
of the black rings  embedded in the asymptotically flat spacetime.

Thus we believe it worthwhile to revisit the identification of the charges
directly in five dimensions, with local five-dimensional Lorentz 
symmetry intact. It poses two related  problems because of the presence
of the Chern-Simons interaction in the Lagrangian.
One is that, in the presence of the Chern-Simons interaction,
the equation of motion of the gauge field is given by
\begin{equation}
d\star F= F\wedge F, \label{eq:CSschematic}
\end{equation} which 
means that the topological density of the gauge field itself
  becomes the source of electric charge. 
To put it  differently,   the attractor mechanism for the black rings
\cite{KLattractor} determines the scalar vevs
at the near-horizon region
via the magnetic dipole charges only, and the information about
the electric charges seems to be lost. 
Then the electric charge of a black ring seems to be diffusely distributed
throughout the spacetime.   Eq.\eqref{eq:CSschematic} can be rewritten
in the form \begin{equation}
d(\star F-A\wedge F)=0,
\end{equation} then $\int_\Sigma (*F-A\wedge F)$ is independent of $\Sigma$.
This integral is called the Page charge.
Similar analysis can be done for angular momenta, and 
Suryanarayana and Wapler \cite{SW} obtained a nice formula for them 
using the Noether charge of Wald.

There is a second problem remaining for black rings,
which stems from the fact that
$A$ is not a well-defined one-form there because
of the presence of the magnetic dipole. It makes
$\int_\Sigma (\star F-A\wedge F)$ ill-defined, because
in the integral all the forms are to be well-defined.
The same can be said for the angular momenta.
The aim of this paper is then to show how this second problem
can be overcome, and to see how the near-horizon region of a
black ring encodes its charges measured at the asymptotic infinity. 

In Section~\ref{sec:uncorrected},
we  use elementary methods to convert the
integral at the asymptotic infinity to the one at the horizon. 
We apply our formalism to the supersymmetric black ring and check that it
correctly reproduces known values for the conserved charges.
We will show how the gauge non-invariance of $\int A\wedge F$ 
can be solved by using two coordinate patches and a compensating term
along the boundary of the patches.
Then in Section \ref{sec:application}
we will  see that our viewpoint helps in
identifying the relation of the charges under the Kaluza-Klein reduction
along $S^1$. We will see that the change in the charges
under a large gauge transformation in five dimensions maps 
to the Witten effect on  dyons \cite{Witten} in four dimensions. 
Proposals in the literature \cite{microring1,microring2,taub1,taub2,taub3}
will be found  equivalent under the transformation.
We conclude with a summary in Section \ref{sec:conclusion}.
In Appendix~\ref{sec:ringgeometry} the geometry of the
concentric rings is briefly reviewed.

\section{Near-Horizon Data and Conserved Charges}
\label{sec:uncorrected}

To emphasize essential
physical ideas, we discuss the problem first for the minimal
supergravity in five dimensions.  Later in this section
we will apply the technique to the case with vector multiplets.
The bosonic part of the Lagrangian of the minimal 
supergravity theory is  \begin{equation}
S=\frac{1}{8\pi G}\int\left( \frac12\star R - F\wedge \star F - \frac{4}{3\sqrt{3}}
A\wedge F\wedge F\right).\label{eq:minimalaction}
\end{equation} Our metric is mostly plus,
and $R_{\mu\nu}$ is defined to be positive for spheres.
We define the Hodge star operator for an $n$-form as
\begin{eqnarray}
\star \left(dx^{\mu_0} \wedge \cdots \wedge dx^{\mu_{n-1}} \right)
= \frac{\sqrt{-g}}{(5-n)!} \epsilon^{\mu_0 \cdots \mu_{n-1}}{}_{\mu_{n} \cdots \mu_4} dx^{\mu_{n}} 
\wedge \cdots \wedge dx^{\mu_4}\;.
\end{eqnarray}
with the Levi-Civita symbol $\epsilon_{01234} = +1$ and $\epsilon^{01234} =  -1$ defined in local Lorentz coordinates.
The equations of motion  are
\begin{align}
R_{\mu\nu}&= - \frac{1}{3}g_{\mu\nu} F_{\rho\sigma} F^{\rho\sigma}
 + 2 F_{\mu\rho}F_{\nu}{}^{\rho},\label{eq:metricEOM}\\
d\star F&=-\frac2{\sqrt{3}}F\wedge F.\label{eq:gaugeEOM}
\end{align}

\subsection{Electric charges}
\label{subsec:uncorrected-ec}

From the equation of motion of the gauge field \eqref{eq:gaugeEOM}, 
we see that $F\wedge F$ is the electric current
for the  charge $\int\star F$. Thus, the charge
is distributed diffusely in the spacetime as was emphasized $\mbox{e.g.}$ in \cite{gauntlett-myers-townsend}.
However, the equation \eqref{eq:gaugeEOM} 
can also be cast in the form \begin{equation}
d\left(\star F+\frac2{\sqrt3} A\wedge F\right)=0.
\end{equation}  At the asymptotic infinity, $A\wedge F$ decays sufficiently rapidly,
so that  we have \begin{equation}
\int_\infty \star F=
\int_\infty \left(\star F+\frac2{\sqrt{3}} A\wedge F\right)
=\int_\Sigma \left(\star F+\frac2{\sqrt{3}} A\wedge F\right),
\label{eq:Page}
\end{equation} where the subscript $\infty$ indicates that the integral
is taken at $S^3$ at the asymptotic infinity, 
and $\Sigma$ is an arbitrary three-cycle
surrounding the black object.
 Thus we can think of the electric charge as the integral of the quantity
inside the bracket, which is called the Page charge.

One problem about the Page charge is that, even in the case where $A$ is a 
globally defined one-form, it changes its value under a large gauge transformation.
It is completely analogous to the fact that $\int_C A$ for an uncontractible circle $C$ 
is only defined up to an integral multiple of $2\pi$ under a large gauge transformation.
Indeed, let us parametrize $C$ by $0\le \theta\le 2\pi$
and we perform a gauge transformation by $g(\theta)\in U(1)$,
i.e. we change $A$ to $A+i^{-1}g^{-1}dg$.
Such a continuous $g(\theta)$ can be written as $g(\theta)=\exp(i\phi(\theta))$.
Then $\int_C A$ changes by $\int_C d \phi(\theta) = \phi(2\pi)-\phi(0)$, which can jump by
a multiple of $2\pi$.  Thus, $\int_C A$ is invariant under a small gauge transformation
$\phi(0)=\phi(2\pi)$ but is not under a large gauge transformation $\phi(0)\ne \phi(2\pi)$.
Exactly the same analysis can be done for $\int_\Sigma A\wedge F$,
and it changes under a large gauge transformation along $C$ if 
$\Sigma$ contains intersecting one-cycle $C$ and two-cycle $S$ and $\int_S F\ne 0$.
However, this non-invariance under a large gauge transformation poses
no problem if $\Sigma$ is at the asymptotic infinity of the flat space,
because we usually demand that $A$ should decay sufficiently rapidly there, which
removes the freedom to do a large gauge transformation.

These facts are well-known, and have been utilized
previously $\mbox{e.g.}$ in \cite{Morales}.
It is the manifestation
of the fact that there are several notions of electric charges
in the presence of Chern-Simons interactions, as clearly discussed by
Marolf in Ref.~\citen{Marolf}.
One is the Maxwell charge which is gauge-invariant but not conserved,
and another is the  charge which is conserved but not gauge-invariant.
In our case $\int \star F$ is the Maxwell charge and $\int(\star F+(2/\sqrt{3})A\wedge F)$
is the Page charge. 
Yet another notion of the  charge  is the quantity which generates
the symmetry in the Hamiltonian framework,
which can be constructed using  Noether's theorem  and its generalization
by the work of Wald and collaborators
\cite{LeeWald,WaldIdent,Wald,IyerWald}.
The charge  thus constructed  is called the Noether charge, 
and in our case it agrees with the Page charge.

Unfortunately, the manipulation above cannot be directly applied to the black rings
with dipole charges. It is because
$A$ is not a globally well-defined one-form, and  the integrals
are not even well-defined.  The way out is to 
generalize the definition of $\int_C A\wedge F$
to the case $A$ is a $U(1)$ gauge field defined using two coordinate patches,
so that \begin{equation}
\int_B F\wedge F = \lquote \int_{\partial B} A\wedge F\rquote 
\label{eq:PagePartialIntegration}
\end{equation} holds. Then the manipulation \eqref{eq:Page}  makes sense. 
The essential idea is to introduce a term localized in the boundary of the patches
which compensates the gauge variation.  Copsey and Horowitz \cite{CH} used
similar subtlety associated to the gauge transformation between patches
to study how the magnetic dipole enters in the first law of the black rings.
%The electric charge at the asymptotic infinity can also be measured as the Page
%charge at the horizon.  From the attractor mechanism, there is no electric field
%near the horizon of the supersymmetric black rings. 
%The equation \eqref{eq:Page} tells us that
%the asymptotic electric charges are instead encoded in the Wilson lines
%$A^I$ along the horizon.  

\FIGURE{
\centerline{\includegraphics[width=.6\textwidth]{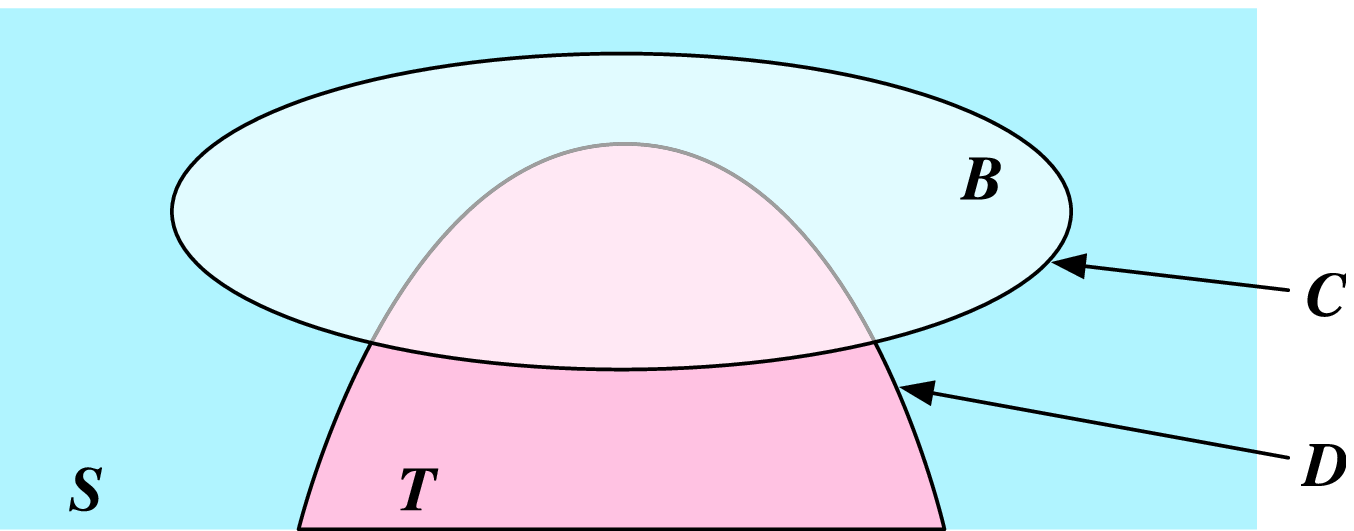}}
\caption{Coordinate patches used to define $\int_C A\wedge F$ consistently 
without ambiguity.\label{fig:AFfigure}}}

Let us assume the whole spacetime is covered by two coordinate patches,
$S$ and $T$, see Figure \ref{fig:AFfigure}. 
We denote the boundary
of two regions by $D=\partial S=-\partial T$.  The gauge field $A$
is represented as   well-defined one-forms $A_S$ and $A_T$ on the patches
$S$ and $T$, respectively.  These two are related by 
a gauge transformation, $A_S=A_T+\beta$ with $d\beta=0$ on the boundary  $D$.
Suppose the region $B$ has the boundary $C=\partial B$.
Then we have 
\begin{eqnarray}
\int_B F\wedge F&=&\int_{B\cap S} F\wedge F+\int_{B\cap T} F\wedge F\\
&=&\int_{C\cap S+ D\cap B} A_S\wedge F+\int_{C\cap T-D\cap B}
A_T\wedge F\\
&=&(\int_{C\cap S} A_S\wedge F+\int_{C\cap T} A_T\wedge F)
+\int_{D\cap B}(A_S\wedge F-A_T\wedge F)\\
&=&(\int_{C\cap S} A_S\wedge F+\int_{C\cap T} A_T\wedge F)
+\int_{C\cap D} A_S\wedge \beta.
\end{eqnarray}

Now we define the symbol $\int_M A\wedge F$ for a three-cycle $M$
to mean \begin{equation}
\lquote\int_M A\wedge F\rquote
\equiv
\int_{M\cap S} A_S\wedge F+\int_{M\cap T} A_T\wedge F+\int_{D\cap M}A_S\wedge \beta ,\label{eq:PageDef}
\end{equation}  then the relation \eqref{eq:PagePartialIntegration} holds as is. 
The important point here is that we need a  term 
$\int_{D\cap M} A_S \wedge \beta$ which compensates the gauge variation
localized at the boundary of the coordinate patches.

  One immediate concern might be
the gauge invariance of the definition \eqref{eq:PageDef}, but 
it is guaranteed for $C=\partial B$ 
from the very fact the relation \eqref{eq:PagePartialIntegration} holds.
It is because its left hand side is obviously gauge invariant.
For illustration, consider the case $\partial B= C_1- C_2$.
The Page charges measured at $C_1$, $C_2$ themselves 
are affected by a large gauge transformation, but their difference is not.
When one takes $C_1$ as the asymptotic infinity, it is conventional
to set the gauge potential to be zero there, thus fixing the gauge freedom.
Then the Page charge at the cycle $C_2$  is defined without ambiguity.

In the following, we drop the quotation marks around the generalized integral
``$\int A\wedge F$~''. We believe it does not cause any confusion.

\subsection{Angular momenta}
\label{subsec:uncorrected-am}
The technique similar to the one  we used for electric charges
can be applied to the angular momenta, and we can obtain a 
formula which expresses them
by the integral at the horizon.  There is a general formalism,
developed by Lee and Wald \cite{LeeWald},
which constructs the appropriate integrand from a given arbitrary generally-covariant
Lagrangian,
and the expression for the angular momenta was obtained in \cite{SW,Rogatko}.
Instead, here we will construct a suitable quantity
in a more down-to-earth and  direct method. We will see that the integrand
contains the gauge field $A$ without the exterior derivative, and that 
it is ill-defined in the presence of magnetic dipole. 
We will use the technique developed in the last section
to make it well-defined.

Firstly, the angular momentum corresponding to the axial Killing vector $\xi$
can be measured at the asymptotic infinity by Komar's formula \begin{equation}
J_\xi = - \frac{1}{16\pi G}\int_\infty \star \nabla \xi,
\end{equation} where $\nabla \xi$ is an abbreviation for 
the two-form $\nabla_\mu \xi_\nu dx^\mu \wedge dx^\nu = d\xi$. 
Using the Killing identity,
the divergence of the integrand is given by \begin{equation}
d\star \nabla\xi = 2\star  R_{\mu\nu} \xi^\mu dx^\nu ,
\end{equation}  which vanishes in the pure gravity. 
Thus, the angular momentum of a black object of the pure gravity theory
can be measured by
$\int_S \star \nabla\xi$ for any surface $S$ which surrounds the object.

Let us analyze  our case, where the equations
of motion are given by \eqref{eq:metricEOM} and \eqref{eq:gaugeEOM}.
We need to introduce some notations:
$\Lie_\xi$ denotes the Lie derivative along the vector field $\xi$,
$\iota_\xi \omega$ denotes the interior product of a vector $\xi$ 
to a differential form $\omega$, i.e.~the contraction
of the index of $\xi$ to the first index of $\omega$.
 Then $\Lie_\xi = d\iota_\xi + \iota_\xi d$
when it acts on the forms.   For a vector $\xi$ and a one-form $A$,
we abbreviate $\iota_\xi A$ as $(\xi\cdot A)$.

We will take
the gauge where gauge potentials are invariant under  the
axial isometry $\Lie_\xi A=0$.  
It can be achieved by 
averaging over the orbit of the isometry $\xi$.
We furthermore assume that every chain or cycle we use
is invariant under the isometry $\xi$, then any term of the form
$\iota_{\xi}(\cdots)$ vanishes upon integration on such a chain or cycle.  

Under these assumptions, the difference of the integral of $\star \nabla \xi$
at the asymptotic infinity and at $C$ is evaluated with the help of
the Einstein equation \eqref{eq:metricEOM} to be \begin{equation}
\int_\infty\star \nabla\xi-\int_C\star \nabla\xi = 2 \int_B \star R_{\mu\nu}\xi^\mu
dx^\nu
= 4 \int_B (\iota_\xi F) \wedge \star F\label{eq:tochu}
\end{equation}where $B$ is a hypersurface connecting the asymptotic infinity and $C$.
We dropped the term $\int\iota_\xi(\star F^2)$ because it vanishes upon integration.

The right hand side can be partially-integrated using
the following relations: one is 
\begin{equation}
d\left[ \star (\xi\cdot A) F\right] = - (\iota_\xi F) \wedge\star  F
 - (\xi\cdot  A) \frac2{\sqrt{3}} F\wedge F
\end{equation} and another is
 \begin{align}
d\left[ (\xi\cdot  A) A\wedge F\right]
&=  (\xi\cdot  A) F\wedge F - ( \iota_\xi F)  \wedge A\wedge F \\
&=  \frac32 (\xi\cdot  A)F\wedge F - \frac12\iota_\xi (A\wedge F\wedge F)
\end{align} of which the last term vanishes upon integration.
Thus we have \begin{equation}
dX_\xi[A]=-(\iota_\xi F)\wedge \star F \label{eq:boo}
\end{equation} modulo the term of the form $\iota_\xi(\cdots)$,
where  \begin{equation}
X_\xi[A]\equiv\star (\xi\cdot  A) F +\frac4{3\sqrt{3}} (\xi\cdot A) A\wedge F.\label{eq:X}
\end{equation}
$X_\xi[A]$ is not a globally well-defined form.  Thus, to perform
the partial integration of 
the right hand side of \eqref{eq:boo},  compensating terms along
the boundary of the coordinate patches need to be introduced,
just as we did in the previous section in the analysis of the Page charge.

Let  $S$ and $T$ be two coordinate patches,
$D=\partial S=-\partial T$ be their common boundary, 
and $A_S=A_T+\beta$ as before.
Let us call the correction term $Y_\xi[\beta,A_S]$ and we define 
\begin{equation}
\int_M  X_\xi[A]
\equiv  \int_{M\cap S} X_\xi[A_S]+\int_{M\cap T} X_\xi[A_T]
+\int_{M\cap D} Y_\xi[\beta,A_T].
\end{equation} We demand that it satisfies \begin{equation}
\int_{\partial B} X_\xi[A]=\int_B (\iota_\xi F)\wedge \star F.
\end{equation}
Then $Y[\beta,A]$ should solve 
\begin{align}
dY_\xi[\beta,A_T]&=X_\xi[A_S]- X_\xi[A_T].
\end{align}  The right hand side is automatically closed since
$dX_\xi[A]$ is gauge invariant. Thus the equation above should have a solution
if there is no cohomological obstruction.
Indeed, substituting \eqref{eq:X} in the above equation, we get \begin{equation}
Y_\xi[\beta,A_T]=(\xi\cdot \beta)Z 
-\frac{2}{3\sqrt3}\bigl[2(\xi\cdot \beta)\beta\wedge A_T+(\xi\cdot A_T)\beta\wedge A_T\bigr]
\end{equation} modulo $\iota_\xi(\cdots)$,
where $dZ$ should satisfy\begin{equation}
dZ=\star F+\frac2{\sqrt3}A_T\wedge F,\label{eq:Z}
\end{equation}the right hand side of which is closed using the equation
of motion \eqref{eq:gaugeEOM}.
Unfortunately there seems to be no
general way to write $Z$ as a functional of $A$ and $\beta$. 
We need to choose $Z$ by hand for each  on-shell configuration.
With these preparation, we can finally  integrate
the right hand side of \eqref{eq:tochu} partially 
and conclude that \begin{equation}
\int_C \left(\star \nabla\xi +4 X_\xi[A]\right).
\end{equation}  is independent under continuous deformation of $C$.

Taking $C$ to be the 3-sphere at the asymptotic infinity,
the terms $X[A]$ vanish too fast to contribute to the integral.
Then,  the integral above is proportional to the Komar integral
at the asymptotic infinity.
Thus we arrive at the formula
\begin{equation}
J_\xi = - \frac1{16\pi G}\int_{\Sigma}\left(\star \nabla \xi + 4
\star (\xi\cdot  A) F + \frac{16}{3\sqrt{3}} (\xi\cdot A) A\wedge F\right),
\label{eq:Jformula}
\end{equation}  
where $\Sigma$ is any surface enclosing the black object.
 The right hand side is precisely the
Noether charge of Wald as constructed in \cite{SW,Rogatko}.

The contribution $\int \star \nabla\xi$ to the angular momentum
is gauge invariant but is not conserved.  It is expected, since
the matter energy-momentum tensor carries the angular momentum.
The rest of the terms in \eqref{eq:Jformula}
was obtained by the partial integral of the
contribution from the matter energy-momentum tensor, and can also
be obtained by constructing the Noether charge. The price we paid is
that it is now not invariant under a gauge transformation. 

\subsection{Example 1: the black ring}
Let us check our formulae against known examples.  First we consider the
celebrated supersymmetric black ring in five dimensions \cite{ring}.
\subsubsection{Geometry}
\label{sec:ringring}
It has been known \cite{allsolution} that
any supersymmetric solution of the minimal supergravity
in the asymptotically
flat $\bR^{1,4}$ can be written 
 in the form
\begin{equation}
ds^2=-f^2 (dt+\omega)^2 + f^{-1}ds^2(\bR^4)\label{eq:ringmetric}
\end{equation} where $f$ and $\omega$ is a function and a one-form
on $\bR^4$, respectively. 
For the supersymmetric black ring \cite{ring},
we use a coordinate system adopted for a ring of radius $R$
in the $\bR^4$ given by \begin{equation}
ds^2(\bR^4)=\frac{R^2}{(x-y)^2}\left[\frac{dx^2}{1-x^2}+(1-x^2)d\phi_1^2
+\frac{dy^2}{y^2-1}+(y^2-1)d\phi_2^2
\right]\label{eq:flat}
\end{equation}
with the ranges $-1\le x \le 1$, $-\infty < y \le -1$ and $0\le \phi_{1,2}<2\pi$.\footnote{
We fix the orientations so that $\int_{\Sigma} dx \wedge d\phi_1 \wedge d\phi_2 > 0$ and 
$\int_{S^2} dx \wedge d\phi_1 < 0$ for $S^2$ surrounding the ring.}
$\phi_1,\phi_2$ were denoted by $\phi,\psi$ in Ref.~\citen{ring}. 

The solution for the single black ring is parametrized by
the radius $R$ in the base $\bR^4$ above, and two extra parameter $q$ and $Q$.
More details can be found in Appendix~\ref{sec:ringgeometry}.
$q$ controls the magnetic dipole through $S^2$ surrounding the ring,
\begin{equation}
\frac{1}{2\pi} \int_{S^2} F = \frac{\sqrt{3}}2 q.
\end{equation}
Conserved charges measured at the asymptotic infinity  are as follows: \begin{align}
\mathbf{Q}=\frac{1}{4\pi G}\int_\infty \star F &= 
	\frac{\sqrt{3}\pi}{2G} Q,\label{eq:echarge}\\
J_1= - \frac{1}{16\pi G}\int_\infty \star  \nabla\xi_1 &= 
	\frac{\pi}{8G}q(3Q-q^2), \label{eq:Jphi}\\
J_2= - \frac{1}{16\pi G}\int_\infty \star  \nabla\xi_2 &= 
	\frac{\pi}{8G}q(6R^2+3Q-q^2) \label{eq:Jpsi0}
\end{align} where $\xi_{1}$, $\xi_{2}$ are the vector fields
$\partial_{\phi_1}$, $\partial_{\phi_2}$ respectively.

There is a magnetic flux through $S^2$ surrounding the ring,
so we need to introduce two patches $S,T$. 
We choose $S$ to cover the region $x < 1-\epsilon$
and $T$ to cover $1-\epsilon < x < 1$, with infinitesimal $\epsilon$.
The boundary $D$ is at $x=\epsilon$ and parametrized by
$0\le \phi_1,\phi_2 <2\pi$.  
We choose the gauge transformation between the two patches to be \begin{equation}
A_T=A_S+\frac{\sqrt{3}}{2} qd\phi_1\label{eq:At}
\end{equation} which is chosen to make $A_T$ smooth at the origin of $\bR^4$.

The horizon is located at $y\to -\infty$ and has the topology $S^1\times S^2$.
The gauge potential near the horizon is \begin{align}
A_S&=-\frac{\sqrt{3}}{4}\left(q+\frac Qq \right) d\psi
-\frac{\sqrt{3}}{4}q(x+1)d\chi,\label{eq:nearhorizongauge}
\end{align} while the geometry near the horizon is given as \begin{equation}
ds^2=2dv d r + \frac{4\ell}{q}r dv d\psi + \ell^2 d\psi{}^2+\frac{q^2}{4}(d\theta^2+
\sin^2\theta d\chi^2)\label{eq:nearhorizonmetric}
\end{equation} where $r=r(y)$ is chosen so that $r\to 0$ corresponds to
$y\to-\infty$, $v$ is a combination of $t$ and $y$, $x=\cos\theta$, 
$\psi=\phi_2+C_1/r+C_0$ for
suitably chosen $C_{0,1}$,  $\chi=\phi_1-\phi_2$, and \begin{equation}
\ell^2=3\left(\frac{(Q-q^2)^2}{4q^2}-R^2\right).
\end{equation}
 It is a direct product 
of an extremal Ba\~nados-Teitelboim-Zanelli (BTZ)
black hole with horizon length $2\pi \ell$ and curvature radius $q$
and of a round two-sphere with radius $q/2$.

 $\ell$ is a more physical quantity 
characterizing the ring than $R$ is, so it is preferable to express 
$J_2$, \eqref{eq:Jpsi0}, using $\ell$ in the form \begin{equation}
J_2= \frac{\pi}{8G} q\left[ -2\ell^2+\frac{3Q^2}{2q^2}-\frac{q^2}2\right].
\label{eq:Jpsi}
\end{equation}Our objective is to reproduce the conserved charges,
\eqref{eq:echarge}, \eqref{eq:Jphi} and \eqref{eq:Jpsi}, purely from
the near-horizon data, \eqref{eq:nearhorizongauge} and \eqref{eq:nearhorizonmetric}.

\subsubsection{Electric charge}
We use the formula \eqref{eq:Page} to get the electric charge.
Using the form of the gauge field near the horizon \eqref{eq:At}
and \eqref{eq:nearhorizongauge},
%and $\beta=A_S-A_T=-\frac{\sqrt{3}}{2} qd\phi_1$, 
we obtain
\begin{align}
\mathbf{Q}&=
\frac{1}{4\pi G} \frac{2}{\sqrt{3}}\int_\Sigma A\wedge F \nonumber \\
&= \frac{1}{4\pi G} \frac{2}{\sqrt{3}}
\left(\int_{S\cap \Sigma} A_S\wedge F + \int_{D\cap \Sigma} A_S\wedge \beta\right)
\nonumber\\
&= \frac{\sqrt{3}\pi}{2G} (\frac{Q+q^2}2+\frac{Q-q^2}2)=\frac{\sqrt{3}\pi}{2G} Q,
\end{align}which correctly reproduces the charge measured at the asymptotic infinity.
Vanishing of $\int_\Sigma \star F$ at the horizon means that all the Maxwell
charge of the system is carried outside of the horizon
in the form of $\int F\wedge F$, while all of the Page charge is
still inside the horizon.

One important fact behind the gauge invariance of the calculation above is that 
the integral $\int A_S$ along the $\psi'$ direction is not just defined mod
integer, but is well-defined as a real number. It is because the circle along
$\psi$, which is not contractible  in the near-horizon region,
becomes contractible in the full geometry.

\subsubsection{Angular momenta}
The integral of the right hand side of \eqref{eq:Z} can be made arbitrarily small
by choosing very small $\epsilon$, so that we can forget  the
complication coming from the choice of $Z$. 
Then for $\xi_1=\partial_{\phi_1}=\partial_{\chi}$,  we have \begin{align}
J_1&=\frac{1}{16\pi G}\left[
-\int_{-1<x<1-\epsilon} \frac{16}{3\sqrt3}(\xi\cdot A_S)A_S\wedge F
+\int_{x=1-\epsilon} \frac{16}{3\sqrt3}(\xi\cdot \beta)\beta\wedge A_T
\right]\nonumber \\
&=\frac{1}{16\pi G}\frac{16}{3\sqrt3}(2\pi)^2\left(\frac{\sqrt3}2\right)^3
\left[\frac14(q^3+qQ) +\frac12 (-q^3+qQ)\right]\nonumber\\
&=\frac\pi{8 G} q(3Q-q^2),
\end{align}  reproducing \eqref{eq:Jphi}.

For $\xi_{\psi}=\partial_{\psi}=\partial_{\phi_1}+\partial_{\phi_2}$, 
we have a contribution from
$\int \star \nabla \xi_\psi = 4\pi^2 q\ell^2$.
Adding contribution from $X[A]$, we obtain \begin{equation}
J_{\psi}=\frac{\pi}{8 G}\left( -2q\ell^2 -\frac{q^3}2+3qQ+\frac{3Q^2}{2q}  \right)
\label{eq:JJJ}
\end{equation} which matches with
$J_{1}+J_{2}$, see \eqref{eq:Jphi} and \eqref{eq:Jpsi}.

The second and the third terms in the expression above are obtained by
the partial integration of the contribution from the angular part of the
energy-momentum tensor of the gauge field. In this sense,
a part of the angular momentum is carried outside of the horizon
and the part proportional to $\ell^2$ is carried inside the horizon.
However, the Noether charge of the black ring
resides  purely inside of the horizon.

\subsection{Example 2: concentric black rings}
The concentric black-ring solution constructed in 
Ref.~\citen{Concentric} is a superposition of the single black ring
we discussed in the last subsection. We focus on the case where all
the rings lie on a plane in the base $\bR^4$.
For the superposition of $N$ rings, the full geometry 
is parametrized by $3N$ parameters $q_i$, $Q_i$ and $R_i$,
$(i=1,\ldots,N)$.
$q_i$ is the dipole charge and $R_i$ is the radius in the base $\bR^4$
of the $i$-th ring.
For more details, see Appendix~\ref{sec:ringgeometry}.
We order the rings so that $R_1<R_2<\cdots<R_N$.
The conserved charges measured at infinity
are known to be
\begin{eqnarray}
{\mathbf Q} &=&\frac{\sqrt3 \pi}{2G} \left[\sum_{i=1}^N \left(Q_i - q_i^2 
\right) + \sumq^2 \right], \label{eq:conc_charge} \\
J_{1} &=& \frac{\pi}{8G} \left[2 \sumq ^3 
+ 3 \sumq  \sum_{j=1}^N (Q_j - q_j^2) \right],
\label{eq:conc_phi1}\\
J_{2} &=& \frac{\pi}{8G} \left[2 \sumq ^3 
+ 3 \sumq  \sum_{j=1}^N (Q_j - q_j^2) + 6 
\sum_{i=1}^N q_i R_i^2 \right] \label{eq:conc_phi2}
\end{eqnarray} 
where $\sumq$ is an abbreviation  for the sum of the magnetic charges, 
i.e.~$s=\sum_{i=1}^N q_i$.
Our aim is to reproduce these results from the near-horizon data.

The near-horizon metric of $i$-th ring has the form 
\eqref{eq:nearhorizonmetric} with $q$, $Q$, $R$ replaced with $q_i$, $Q_i$ and 
$R_i$, respectively.
The horizon radius $\ell_i$ is given by \begin{equation}
\ell_i^2=3\left(\frac{(Q_i-q_i^2)^2}{4q_i}-R_i^2\right).
\end{equation} 

Since each ring has a magnetic dipole charge, we introduce 
coordinate patches $S$ and $T_i$ so that the gauge field is non-singular in 
each patch. Let $T_i$ be the patch covering the region between $(i-1)$-th and 
$i$-th ring and $S$ be a patch covering the outer region. More precisely,
we introduce the ring coordinate \eqref{eq:flat} for each of the ring,
 and choose
$S$ to cover $-1 + \epsilon <  x_i <  1-\epsilon$ for each ring while $T_i$ to cover
$1 - \epsilon <  x_i <  1$ for the $i$-th ring and $-1 <  x_{i-1} <  -1 + \epsilon$ 
for the $(i-1)$-th ring. Then, near the 
$i$-th horizon the gauge field on $S$ is given by
\begin{eqnarray}
A_S=-\frac{\sqrt3}{4}\left[\left(\frac{Q_i}{q_i}-q_i+2\sumq\right)d\psi+\left(q_i(1+x)+ 2\sum_{j=i+1}^N q_j\right)d\chi\right].
\end{eqnarray}
Its $\psi$ component is determined in 
Appendix~\ref{sec:ringgeometry}, while 
 the coefficient for $d\chi$ is determined so that the field strength is 
reproduced, the gauge field is non-singular except for $x= \pm 1$ for the 1st 
to $(N-1)$-th rings and non-singular except for $x=-1$ for the $N$-th ring. 
The gauge field on $T_i$ is given by
\begin{eqnarray}
A_{T_i}=A_S+\frac{\sqrt3}2\sum_{j=i}^N q_j d\phi_1.
\end{eqnarray}

The electric charge is given by using \eqref{eq:Page} and $\beta_i = 
A_S - A_{T_i}= - \frac{\sqrt3}{2} \sum_{j=i}^N q_j d \phi_1$ as
\begin{eqnarray}
{\mathbf Q} &=& 
 \frac{1}{4 \pi G} \frac{2}{\sqrt3} \sum_{i=1}^N \int_{\Sigma_i \cap S} A_S 
\wedge F+ \frac1{4 \pi G} \frac2{\sqrt3} 
\sum_{i=1}^N \left( \int_{\Sigma_i \cap \partial S} 
A_S \wedge \beta_i + \int_{\Sigma_{i-1} \cap \partial S} 
A_S \wedge \beta_i \right)
\nonumber\\
{}&=& \frac{\sqrt3 \pi}{4 G} 
\sum_{i=1}^N \left[ \left(Q_i - q_i^2 \right) + 2 s q_i \right] + \frac{\sqrt3 \pi}{4 G} 
\sum_{i=1}^N \left(Q_i - q_i^2 \right) \nonumber \\
{}&=& \frac{\sqrt3 \pi}{2G} \left[ \sum_{i=1}^N \left(Q_i - q_i^2 \right) 
+ \sumq^2\right].
\end{eqnarray}
This correctly reproduces the known result \eqref{eq:conc_charge}.

Let us move onto the evaluation of the angular momenta. Note that for 
certain configurations of charges, the concentric black rings develop 
singularities on the rotation axes. While the condition for the absence of 
singularities has not been known fully, it was pointed out in 
Ref.~\citen{Concentric} that there is no singularity on the rotation axes if 
all
\begin{eqnarray}
\Lambda_i = \frac{Q_i - q_i^2}{q_i}
\label{eq:non_singular}
\end{eqnarray}
are equal. We will show that we can obtain the correct angular momenta if this 
condition is satisfied.

The angular momentum associated with 
$\xi_1 = \partial_{\phi_1}= \partial_{\chi}$ is given by
\begin{eqnarray}
J_1 &=& - \frac1{16 \pi G} \frac{16}{3\sqrt3} \sum_{i=1}^N 
\int_{\Sigma_i \cap S} 
(\xi_1 \cdot A_S) A_S \wedge F \nonumber \\
{}&& 
- \frac{1}{16 \pi G}\frac{16}{3\sqrt3} \sum_{i=1}^N 
\left( \int_{\Sigma_i \cap \partial T_i} + \int_{\Sigma_{i-1} \cap \partial 
T_i} \right) (\xi_1 \cdot \beta_{i})\beta_{i} \wedge A_{T_i} .
\label{eq:conc_nhphi1}
\end{eqnarray}
After summing up terms, we have
\begin{equation}
J_1 = \frac{\pi}{8G} \left[ 2 s^3 
+ 6 \sum_{i=1}^N (Q_i - q_i^2) \sum_{j=i+1}^N q_j 
+ 3 \sum_{i=1}^N (q_i(Q_i - q_i^2))  \right] .
\end{equation}
If the condition \eqref{eq:non_singular} is satisfied, $J_1$ computed 
above matches \eqref{eq:conc_phi1} and we have
\begin{eqnarray}
J_1 \to \frac{\pi}{8G} \left[ 2 s^3 + 3 \Lambda_i s^2 \right] .
\end{eqnarray}

Finally, let us consider the angular momentum associated with 
$\xi_{\psi} = \partial_{\psi} = \partial_{\phi_1} + \partial_{\phi_2}$. 
In addition to  \eqref{eq:conc_nhphi1} with
$\xi_1$ being replaced by $\xi_\psi$, here we have to 
consider two more contributions. Namely,
\begin{equation}
- \frac1{16 \pi G} \sum_{i=1}^N \int_{\Sigma_i} \star \nabla 
\xi_{\psi} -
\frac1{16 \pi G} \frac8{3 \sqrt3} \sum_{i=1}^N 
\left( \int_{\Sigma_i \cap \partial T_i} + \int_{\Sigma_{i-1} \cap 
\partial T_i} \right)
(\xi_{\psi} \cdot A_{T_i}) \beta_i \wedge A_{T_i}. 
\end{equation}
It is easy to check that the sum of each term is given by
\begin{equation}
J_{\psi} = \frac{\pi}{8G} \left[6\sum_{i=1}^N q_i R_i^2 
+ 4s^3 + 6 s \sum_{i=1}^N (Q_i - q_i^2) \right].
\end{equation}
When evaluated under the condition \eqref{eq:non_singular}, this gives
\begin{eqnarray}
J_{\psi} \to \frac{\pi}{8G} \left[ 6 \sum_{i=1}^N q_i R_i^2 
+ 4 s^3 + 6 \Lambda_i s^2\right]
\end{eqnarray}
and agrees with $J_{\psi}$ given as the sum of \eqref{eq:conc_phi1} and 
\eqref{eq:conc_phi2}.

\subsection{Generalization}
\label{sec:generalization}
It is straightforward to generalize the techniques we developed  so far
to the supergravity theory with $n$ of $U(1)$ vector fields $A^I$, $(I=1,\ldots,n)$. 
There are $(n-1)$ vector multiplets
because the gravity multiplet also contains the graviphoton field
which is a vector field.   The scalars in the vector multiplet are denoted
by $M^I$, which are constrained by the condition \begin{equation}
\cN\equiv c_{IJK} M^IM^JM^K=1.
\end{equation} $c_{IJK}$ is a set of constants.
The action  for the boson fields is given by \begin{equation}
S=\frac1{16\pi G}\int
\left[\star R - a_{IJ}dM^I \wedge \star dM^J - a_{IJ}F^I \wedge \star F^J
- c_{IJK}A^I\wedge F^J \wedge F^K\right]\label{eq:5daction}
\end{equation} where $R$ is the Ricci scalar,
and \begin{equation}
a_{IJ}=-\frac12\left(\cN_{IJ}-\cN_I\cN_J\right).
\end{equation} 
In the last expression, $\cN_I=\partial \cN/\partial M^I$
and $\cN_{IJ}=\partial^2\cN/\partial M^I\partial M^J$.
This is the low-energy action of M-theory compactified on 
a Calabi-Yau manifold $M$ with $n=h_{1,1}(M)$,
and \begin{equation}
6c_{IJK}=\int \omega_I\wedge \omega_J\wedge \omega_K\label{eq:CYtriple}
\end{equation} 
is the triple intersection of integrally-quantized two-forms $\omega_I$ on $M$.
The action for the minimal supergavity \eqref{eq:minimalaction}
is obtained by setting $n=1$, $c_{111}=(2/\sqrt3)^3$, and $a_{11}=2$.

As for the 
calculation of the electric charges, one only needs to put the indices $I,J,K$
to the vector fields and the result is 
\begin{equation}
\mathbf{Q}_I=\frac{1}{8\pi G}\int \left[\star a_{IJ}F^J+\frac32 c_{IJK}A^J \wedge F^K\right].
\label{eq:generalPage}
\end{equation}
As for the angular momenta, there is  extra terms
coming from the energy-momentum tensor of the scalar fields 
in the right hand side of \eqref{eq:tochu}.
Its contribution to the angular momenta vanishes upon integration,
so that the result is 
\begin{equation}
J_\xi= - \frac1{16 \pi G}\int \left[\star \nabla\xi + 
2\star a_{IJ}(\xi\cdot A^I)F^J+2c_{IJK}(\xi\cdot A^I)A^J\wedge F^K\right].\label{eq:generalAng}
\end{equation}

For a more complicated Lagrangian, e.g.~with charged hypermultiplets 
and/or with higher-derivative corrections, it is easier to utilize the general
framework set up by Wald, than to find the partial integral in \eqref{eq:tochu}
by inspection.  The charge constructed by this technique 
has an important property \cite{LeeWald}
that it acts as the Hamiltonian for the corresponding
local symmetry in the Hamiltonian formulation of the theory,
and it reproduces the Page charge and the angular momenta \eqref{eq:generalAng}.
Consequently, the charge as the generator of the symmetry is not the
gauge-invariant Maxwell charge, but the Page charge which depends on a 
large gauge transformation.

The integrands in the expressions above are not well-defined as differential forms
when there are magnetic fluxes, thus it needs to be defined appropriately as we
did in the previous sections.  Generically, we would like to rewrite
the integral of a gauge invariant form $\omega$ on  a region $B$ to 
the integral of $\omega_{(1)}$ satisfying\begin{equation}
d\omega_{(1)}=\omega\label{AA}
\end{equation} on its boundary $\partial B$. 
The problem is that $\omega_{(1)}$ may depend on the gauge.
On two patches $S$ and $T$, it is represented by differential forms
$\omega_{(1)}^{S}$ and $\omega_{(1)}^{T}$
 respectively.  Since $\omega$ is  gauge-invariant,
we have $d\omega_{(1)}^S=d\omega_{(1)}^T$. Thus, if we take a sufficiently
small coordinate patch, we can choose $\omega_{(2)}^{(S,T)}$ such that \begin{equation}
d\omega_{(2)}^{(S,T)}=\omega_{(1)}^S-\omega_{(1)}^T.\label{BB}
\end{equation}
Then one defines the integral of $\omega_{(1)}$ on $C=\partial B$ via \begin{equation}
\int_C \omega_{(1)}\equiv \int_{C\cap S} \omega^S_{(1)}
+\int_{C\cap T} \omega^T_{(1)}
+\int_{C\cap D} \omega^{S,T}_{(2)},\label{eq:generaldef}
\end{equation} where $D=\partial S=-\partial T$.
The equations \eqref{AA}, \eqref{BB} are the so-called descent relation which is
important in the understanding of the anomaly.  It will be interesting to generalize
our analysis to the case where there are more than two patches and 
multiple overlaps among them. Presumably we need to include higher descendants
$\omega_{(n)}^{(S_1,\ldots, S_n)}$ as the correction term at the boundary
of $n$ patches $S_1,\ldots,S_n$ in the definition of the integral \eqref{eq:generaldef}.

\section{Relation to Four-Dimensional Charges}
\label{sec:application}

We have seen how the near-horizon data of the black rings
encode the charges measured at the asymptotic infinity. 
We can also consider rings in the Taub-NUT space \cite{taub1,taub2,taub3}
instead in the five-dimensional Minkowski space.  Then the theory
can also be thought of as a theory in four dimensions, via the 
Kaluza-Klein reduction along $S^1$ of the Taub-NUT space.
It has been established \cite{4d5d} that supersymmetric
solutions for five dimensional supergravity nicely reduces to 
supersymmetric solutions for the corresponding four dimensional theory.

In four dimensions,  there are no problems in defining the charges, 
because the equations of motion
and Bianchi identities yield the relations \begin{equation}
dF^I=0,\qquad d G_I=d(\star (g^{-2}_{IJ})F^J+\theta_{IJ}F^J)=0
\end{equation} where $(g^{-2})_{IJ}$ are the inverse coupling constants
and $\theta_{IJ}$ are the theta angles.
The electric and magnetic charges can be readily obtained by
integrating $G_I$ and $F^I$ over the horizon.
Then it is natural to expect that our formulae for the charges will
yield the four-dimensional ones after the Kaluza-Klein reduction.
One apparent problem is that the Page charges changes
under a large gauge transformation, whereas the four-dimensional
charges are seemingly well-defined as is. We will see that a 
large gauge transformation corresponds to the Witten effect on dyons
in four-dimensions.  
%Most of the observations below seem to be
%known to the experts in the field, but we think it worthwhile to
%discuss it in one place.

\subsection{Mapping of the fields}
First let us recall the well-known mapping of the fields
in four and five dimensions.  
The details can be found e.g.~in \cite{microhigher2,CaiPang,Cardoso,ringrule}.
When we reduce a five-dimensional
$\cN=2$ supergravity with $n$ vector fields along $S^1$,
it results in a four-dimensional $\cN=2$ supergravity  with $(n+1)$ vector fields.
The metrics in respective dimensions are related by \begin{equation}
ds^2_{5d}=e^{2\rho}(d\psi-A^0)^2+e^{-\rho}ds^2_{4d},
\end{equation} where we take the periodicity of $\psi$ to be $2\pi$
so that $e^\rho$ is the five-dimensional radius of the 
Kaluza-Klein circle. The factor in front of the four-dimensional metric
is so chosen that the four-dimensional Einstein-Hilbert term is canonical.

The gauge fields in four and five dimensions are related by \begin{equation}
A^I_{5d}=a^I(d\psi-A^0)+A^I_{4d}\label{eq:gaugeansatz}
\end{equation} where $I=1,\ldots,n$.  It is chosen so that a gauge transformation
of $A^0$ do not affect $A^I_{4d}$.  We need to introduce
coordinate patches when there is a flux for $A^I_{5d}$. We demand that
gauge transformations used between patches should not depend on $\psi$
so that $a^I$ are globally well-defined scalar fields.

Then, by the reduction of the five-dimensional
action \eqref{eq:5daction},
the action of four-dimensional gauge fields is determined to be 
\footnote{We take the following 
conventions in four dimensions: The orientations in 
four and five dimensions are related such that $\int_{5d} dx^0 \wedge dx^1 \wedge 
dx^2 \wedge dx^3 \wedge d \psi =2 \pi \int_{4d} dx^0 \wedge dx^1 \wedge dx^2 \wedge 
dx^3$. The Levi-Civita symbol in four dimensions is 
defined by $\epsilon_{0123} = +1$ and  
$\epsilon^{0123} =-1$ in local Lorentz coordinates.} 
\begin{multline}
\cL= - \left[ \frac12 e^{3\rho}+e^{\rho}a_{IJ} a^I a^J\right]F^0\wedge \star F^0
-c_{IJK} a^I a^Ja^K F^0\wedge F^0\\
+2e^{\rho}a_{IJ}a^I F^0\wedge \star F^J +3c_{IJK}a^Ia^J F^0\wedge F^K\\
-e^\rho a_{IJ}F^I\wedge \star F^J-3c_{IJK} a^I F^J\wedge F^K.\label{eq:4daction}
\end{multline} Partial integrations are necessary to bring the naive Kaluza-Klein
reduction to the form above.
The resulting Lagrangian above follows from the prepotential \begin{equation}
F(X)=\frac{c_{IJK}X^IX^JX^K}{X^0},
\end{equation} if one defines
 special coordinates $z^I=X^I/X^0$  by \begin{equation}
z^I=a^I+i e^\rho M^I. \label{eq:zMrelation}
\end{equation} This relation can 
be checked without the detailed Kaluza-Klein reduction.  Indeed,
the ratio of $a^I$ and $M^I$ in \eqref{eq:zMrelation} can be fixed by
inspecting the mass squared of a hypermultiplet, and the fact
$a^I$ should enter in $z^I$ linearly with unit coefficient is fixed by the monodromy.

\subsection{Mapping of the charges} 
\label{sec:mapping-charges}
In many references including Ref.~\citen{CaiPang,ringrule,SW},
the charge of the black object in five dimensions is defined to be 
the charges in four dimensions after the dimensional reduction
determined from the Lagrangian \eqref{eq:4daction}.
It was motivated partly because the analysis of the charge in five dimensions
was subtle due to the presence of the Chern-Simons interaction,
whereas we studied how we can obtain
the formula for the charges which has five-dimensional general covariance
in Section \ref{sec:uncorrected}. 
Now let us  compare the charges thus defined in four- and five- dimensions.

Firstly, the magnetic charge \begin{equation}
q^0=\frac1{2\pi}\int_C F^0
\end{equation} in four dimensions counts the number of
the Kaluza-Klein monopole inside $C$. It is also called the nut charge.
The other magnetic charges in  four dimensions 
\begin{equation}
q^I=\frac1{2\pi}\int_C F^I
\end{equation} come directly from the dipole charges in five dimensions,
as long as the surface $C$ does not enclose the nut.
When $C$ does contain a nut, the Kaluza-Klein circle is non-trivially
fibered over $C$. Thus, the surface $C$  cannot be lifted to five dimensions.
We will come back to this problem in Section \ref{sec:monodromy-taub}.

The formulae for the electric charges follow from the Lagrangian : 
\begin{align}
Q_I& =\frac1{2\pi}\int\left[ \star 2e^{\rho}a_{IJ}(F^J-a^J F^0)
   + 6 c_{IJK} a^J F^K - 3 c_{IJK} a^Ja^K F^0\right], \label{eq:electric4d}\\
Q_0&=\frac1{2\pi}\int\Bigl[
 \star  e^{3\rho}F^0 -  \star 2e^{\rho}a_{IJ} a^I  (F^J-a^J F^0) +
%\nonumber\\&\qquad\qquad\qquad\qquad
 2c_{IJK}a^Ia^Ja^K F^0  -3 c_{IJK} a^I a^J F^K\Bigr].\label{eq:momentum4d}
\end{align}

It is easy to verify that 
the five-dimensional Page charges \eqref{eq:generalPage}
and the Noether charge 
$J_\psi$ \eqref{eq:generalAng} for the isometry $\partial_\psi$
along the Kaluza-Klein circle are related to the four-dimensional
electric charges via \begin{equation}
Q_I= - \frac{4G}{\pi} \mathbf{Q}_I,\qquad Q_0= - \frac{4G}{\pi} J_\psi.
\end{equation}
 An important point in the calculation is that 
the compensating term on the boundary of the coordinate patches vanishes,
since $a^I$ and $F^J_{4d}$ are globally well-defined.

Thus we see that the four-dimensional charges
are not the reduction of the gauge-invariant Maxwell charges $\int \star F$
or that of the gauge-invariant ``Maxwell-like'' part of the angular momentum,
$\int \star \nabla\xi$.  They are rather the reduction of the Page or the Noether charges,
which change under a large gauge transformation.

\subsection{Reduction and the attractor}
In the literature, the attractor equation is often analyzed after the 
reduction to four dimensions \cite{CaiPang,Cardoso,ringrule},
while the five-dimensional attractor mechanism for the black rings
in \cite{KLattractor} only determines the scalar vacuum expectation
values via the magnetic dipoles. As we saw in the previous sections,
the electric charges at the asymptotic infinity are encoded by the Wilson
lines along the horizon.  We show that how these five-dimensional
consideration reproduces the known attractor solution \cite{deWit,Shmakova}
in four-dimensions.

The five-dimensional metric is characterized by the magnetic charges
$q^I$ through the horizon, and the physical radius of the horizon 
$\ell=e^\rho$ there.   From the attractor mechanism
for the black rings \cite{KLattractor},
the near-horizon geometry is of the form $AdS_3\times S^2$,
and the curvature radii are $q$ and $q/2$ in each factor,
where $q^3=c_{IJK}q^Iq^Jq^K$.
The scalar vevs are fixed to be proportional to the magnetic dipoles,
i.e.~$M^I=q^I/q$.

For the calculation of electric charges the Wilson lines $a^I$
along the horizon are also important. Then we can evaluate the Page charges
and angular momenta on the horizon to obtain \begin{equation}
Q_I=6 c_{IJK}a^J q^K,\qquad Q_0=q\ell^2-3c_{IJK}a^Ia^Jq^K.
\end{equation} We can solve the equations above for $\ell$ and $a^I$
so that we have the formula
for the four-dimensional special coordinates $z^I$ in terms of the charges.
The result is  \begin{equation}
z^I=a^I+ie^\rho M^I= \frac16D^{IJ}Q_I+i \sqrt{\frac{\hat Q_0}{D}} q^I
\end{equation}where \begin{equation}
D_{IJ}= c_{IJK}q^K, \quad D^{IJ}D_{JK}=\delta^I_K
\end{equation} and \begin{equation}
D=q^3=c_{IJK} q^I q^J q^K,\quad \hat Q_0=q\ell^2=Q_0+\frac1{12} D^{IJ} Q_I Q_J.
\label{eq:Qhat}
\end{equation}
It is the well-known solution of the attractor equation in four-dimensions
with $q^0=0$ \cite{deWit,Shmakova}.

Thus, the combination of the attractor mechanism in five dimensions 
and the technique of Page charges yield the attractor mechanism in four dimensions.
The point is that the Wilson lines $a^I$ along the horizon of the black string
carry the information of its electric charges.  Conversely, the Wilson line
at the horizon is determined by the electric charge.  
The horizon length is also determined by the angular momentum.
In this sense, the attractor mechanism for the black rings
also fixes all the relevant near-horizon data by means of the charges,
angular momenta and dipoles.

\subsection{Gauge dependence and monodromy}

Let us now come back to the question of the variation of the Page charges
under large gauge transformations.  The problem is that the integral 
$\int_C A\wedge F$
 depends on the shift $A\to A+\beta$ for $d\beta=0$
if $C$ has a non-contractible loop $\ell$ and  $\int_\ell \beta\ne0$.
In the spacetime which asymptotes to $\bR^{4,1}$,
the large gauge transformation can be fixed by demanding that 
the gauge potential vanishes at the asymptotic infinity.

In the present case of reduction to four dimensions, however, 
the gauge potential along the Kaluza-Klein circle is one of the moduli 
and is not a thing to be fixed. More precisely, if the $\psi$ direction
is non-contractible, a large gauge transformation associated to the
Kaluza-Klein circle corresponds to a shift $a^I\to a^I+t^I$ where $t^I$ are
integers.  In four-dimensional language it is the shift \begin{equation}
z^I\to z^I+t^I,\label{eq:monodromygenerator}
\end{equation}
and the gauge variation of the Page charge translates to the variation
of the electric charge under the transformation \eqref{eq:monodromygenerator}.
It is precisely the Witten effect on dyons \cite{Witten} if one
recalls the fact that the dynamical theta angles of the theory depends on $z^I$.
In the terminology of $\cN=2$ supergravity and special geometry, 
it is called the monodromy transformation associated to the shift \eqref{eq:monodromygenerator}, which acts symplectically
on the charges $(q^I,Q_I)$ and on the projective
special coordinates $(X^I,F_I)$

For the M-theory compactification on the product of $S^1$ and a Calabi-Yau,
electric charges $Q_I$ and $q^I$ correspond to the number of M2-branes
and M5-branes wrapping two-cycles $\Pi^I$ and four-cycles $\Sigma_I$, 
respectively.  
The relation \eqref{eq:CYtriple} translates to 
$6c_{IJK}=\#(\Sigma_I\cap\Sigma_J\cap\Sigma_K)$ in this language.
The gauge fields $A_I$ arise from the Kaluza-Klein
reduction of the M-theory three-form $C$ on $\Pi^I$. 
Thus, the results above imply
that the M2-brane charges transform non-trivially in the presence of M5-branes
under  large gauge transformations of the $C$-field. 

It might sound novel, but it can be clearly seen from the 
point of view of Type IIA string theory on the Calabi-Yau.  
Consider a soliton without D6-brane charge.
There, the D2-brane charge $Q_I$
of the soliton is induced by the world-volume gauge field $F$ on the D4 brane 
wrapped on a four-cycle $\Sigma=q^I\Sigma_I$
through the Chern-Simons coupling \begin{equation}
\int_\Sigma (F+B)\wedge C
\end{equation} where $B$ is the NSNS two-form and $C$ is the RR three-form.
In this description, $a^I$ is given by $\int_{\Pi^I} B$.
The induced brane charge in the presence of the non-zero
$B$-field is an intricate problem in itself, but the end result is that 
the large gauge transformation $B\to B+\omega$ with $\int_{\Pi^I} \omega=t^I$
changes the D2-brane charge of the system by $6c_{IJK}q^I t^J$.  
It will be interesting to derive the same effect from the worldvolume
Lagrangian \cite{M5} of the M5 brane, which is subtle because 
the worldvolume tensor field is self-dual.
The change in the M2-brane charge induce a change in the Kaluza-Klein
momentum carried by the zero-mode on the black strings wrapped on 
$S^1$, so that $Q_0$ also changes \cite{MSW}. The point is that the momentum
carried by non-zero modes, $\hat Q_0$ defined
in \eqref{eq:Qhat}, is a monodromy-invariant quantity.

Before leaving this section, it is worth noticing that
if an M2-brane has the worldvolume $V$, it enters in the equation of motion for
$G=dC$ in the following way: \begin{equation}
d\star G+G\wedge G=\delta_V
\end{equation} where $\delta_V$ is the delta function supported on $V$.
Thus, the quantized M2-brane charge is not the
source of the Maxwell charge.  It is rather the source of the Page charge. 
Essentially the same argument in five dimensions, using the specific decomposition
\eqref{eq:ringmetric}, was made in Ref.~\citen{hairy}.

\subsection{Monodromy and Taub-NUT}
\label{sec:monodromy-taub}
If we use the Taub-NUT space in the dimensional reduction, in other words
if there is a Kaluza-Klein monopole in the system, the Kaluza-Klein circle
shrinks at the nut of the monopole.  As the circle is now contractible,
one might think that one can no longer do a large gauge transformation
and that it is natural to choose $a^I=0$ at the nut. 
Nevertheless, from a four-dimensional
standpoint the monodromy transformation should be always possible.
How can these two points of view be reconciled? 

Firstly, the fact that the five-dimensional spacetime is smooth
at the nut only requires that the gauge field strength is zero there
and that the integral of the gauge potential is an integer.
There should be a patch around the nut in the five-dimensional spacetime 
in which $A^I$ should be smooth, but it is not the patch connected
to the asymptotic region of the Taub-NUT space where $a^I$ is defined.

A similar problem was studied in Ref.~\citen{GHM}. There, it was shown
how the winding number can still be conserved in the background with
the nut, where the circle on which strings are wound degenerates. A crucial
role is played by the normalizable self-dual two-form $\omega$ localized at the nut,
which gives the worldvolume gauge field $A$ of the D6-brane 
realized as the M-theory Kaluza-Klein monopole
via $C=A\wedge \omega$.   
It should enter in the worldvolume Lagrangian in the combination
$dA+B$, and the large gauge transformation affects the contribution from $B$.

Indeed, the Kaluza-Klein ansatz of the gauge fields \eqref{eq:gaugeansatz},
one can make the combined shift \begin{equation}
a^I\to a^I+t^I,\qquad  A^I_{4d}\to A^I_{4d}+t^I A^0
\end{equation} without changing the five-dimensional gauge field strengths.
Therefore, the magnetic charge also transforms as \begin{equation}
q^I \to q^I+t^I q^0.
\end{equation}
The action of the transformation \eqref{eq:monodromygenerator}
on the electric charges  then becomes \begin{align}
Q_I &\to Q_I + 6 c_{IJK} t^J q^K+3 c_{IJK}t^Jt^K q^0,\\
Q_0 &\to Q_0 -  Q_I t^I -3c_{IJK} t^I t^J q^K  -c_{IJK}t^It^Jt^K q^0,
\end{align} which is exactly how
 the projective coordinates \begin{equation}
X^0,\quad X^I, \quad F_I=3c_{IJK}X^JX^K/X^0,\quad F_0=-c_{IJK}X^IX^JX^K/(X^0)^2.
\end{equation} get transformed by the monodromy $a^I\to a^I+t^I$.
It was already noted in Ref.~\citen{taub3} that 
the same symmetry acts on the functions which characterize 
the supersymmetric solution on the Taub-NUT, $(V,K^I,L_I,M)$ in their notation.
The point is that it modifies the five-dimensional Page charges, and hence the 
four-dimensional charges.

If we neglect quantum corrections coming from instantons wrapping
the Kaluza-Klein circle,
it is allowed to do
the monodromy transformation $z^I\to z^I+t^I$
 even with continuous parameters $t^I$.
It maps a solution of the equations of motion
to another, and the electric charges in four-dimensions depends
continuously on the vevs for the  moduli $a^I$ at the asymptotic infinity.
The issue concerning the stability of the solitons can be safely ignored.
In the analyses in Refs.~\citen{taub1,taub2,taub3},
their proposals for the identification of four-dimensional electric
charges $Q_I$ and of five-dimensional ones $\mathbf{Q}_I$
were different from one another. 
%while they all resulted in  correct counting of 
%the microscopic entropy.
The source of the discrepancy in the identification
is now clear after our long discussion.
It can be readily checked that 
the differing  proposals for the identification can be connected
by the monodromy transformation with $t^I = \frac{1}{2}q^I$. 
Namely, the charges in the five-dimensional language are transformed as
\begin{eqnarray}
\frac{4G}{\pi}\mathbf{Q}_I \to \frac{4G}{\pi}\mathbf{Q}_I -3 c_{IJK} q^J q^K,\quad
J_{\psi} \to J_{\psi} - J_{\phi}
\end{eqnarray}
for $Q_0 \gg q^3$ limit.\footnote{We noticed that a small discrepancy
proportional to $c_{IJK}q^Iq^Jq^K$ remains, which is related to the zero-point
energy of the conformal field theory of the black string.  
Its effect on the entropy is subleading
in the large $Q_0$ limit.} Thus they
are equivalent under a large gauge transformation.

The analysis above also answers 
the question raised in Section \ref{sec:mapping-charges}
how the dipole charges in five dimensions are related in the magnetic charges
in four dimensions in the presence of the nut.
It is instructive to consider the case of a black ring in the Taub-NUT space.
From a five-dimensional viewpoint, the dipole charge is not a conserved 
quantity measurable at the asymptotic infinity. Correspondingly, the surface
of the Dirac string necessary  to define the gauge potential 
can be chosen to fill the disc inside the black ring only,
and not to extend to the asymptotic infinity. It was what we did in
Section \ref{sec:ringring} in defining the coordinate patches. 
However,  the gauge transformation required to achieve it necessarily
depends on the $\psi$ coordinate, which is the direction along the Kaluza-Klein
circle. Hence it is not allowed if one carries out the reduction to four dimensions.
In this case, the Dirac string emanating from the black ring 
necessarily extends all the way to the spatial infinity, thus
making the magnetic charge measurable  at the asymptotic infinity.
A related point is that dipole charges enter in the first law of black objects 
because of the existence of two patches \cite{CH}\footnote{
The authors of \cite{CH} used the approach to the first law 
developed in \cite{SudarskyWald}. There is another understanding
of appearance of the dipole charges in the first law \cite{quasilocalring}
if one follows the approach in \cite{quasilocal}. }.
It is easier to understand it after the reduction because now it is a conserved
quantity measurable at the asymptotic infinity.

As a final example to illustrate the subtlety
in the identification of the four- and five-dimensional charges,
let us consider a two-centered Taub-NUT space
with centers $\bx_1$ and $\bx_2$.
There is an $S^2$ between two centers, and one can introduce a 
self-dual magnetic
fluxes $q^I$ through it. 
Although the Chern-Simons interactions put some constraint
on the allowed $q^I$, there is a supersymmetric solution of this form \cite{selfgrav}. 
In this configuration, the Wilson lines $a^I$ 
at $\bx_1$ and $\bx_2$ necessarily differ by the amount
proportional to the flux, and one cannot simultaneously make them zero.
An important consequence is that the magnetic charges $F^I_{4d}$ of the 
nuts at $\bx_2$ and $\bx_2$ necessarily differ, in spite of the fact
that the geometry and the gauge fields in five dimensions 
are completely symmetric under the exchange of $\bx_1$ and $\bx_2$.

\section{Summary}
\label{sec:conclusion}
In this paper, we have first clarified
how the near-horizon data of black objects
encode the conserved charges measured at asymptotic infinity.
Namely, the existence of the Chern-Simons coupling means 
that $F\wedge F$ is a source of electric charges, thus it was necessary
to perform the partial integration once to rewrite the asymptotic electric charge
 by the integral of $A\wedge F$ over the horizon.
Since $F$ has magnetic flux through the horizon, $A\wedge F$ cannot be
naively defined, and we showed how to treat it consistently.
Likewise, we obtained the formula for the angular momenta 
using the near-horizon data.

Then, we saw how our formula for the charges in five dimensions
is related to the four-dimensional formula under Kaluza-Klein reduction.
We studied how the ambiguity coming from large gauge transformations
in five dimensions corresponds to the Witten effect and the associated
monodromy transformation in four dimensions.

It is now straightforward to obtain the correction to the entropy
of the black rings, since we now have the supersymmetric higher-derivative
action \cite{HOT}, the near-horizon geometry \cite{CDKL1,CDKL2,Alishahiha}, 
 and also the formulation developed in this paper
to obtain conserved charges from the near-horizon data alone. 
It would be interesting to see if it matches with the microscopic calculation.

\acknowledgments
YT would like to thank Juan Maldacena, Masaki Shigemori
and Johannes Walcher for discussions.
KH is supported by the Center-of-Excellence (COE) Program ``Nanometer-Scale 
Quantum Physics" conducted by Graduate Course of Solid State Physics and 
Graduate Course of Fundamental Physics at Tokyo Institute of Technology. The 
work of KO is supported by Japan Society for the Promotion of Science (JSPS) 
under the Post-doctoral Research Program. YT is supported by the United States 
DOE Grant DE-FG02-90ER40542.

\appendix

\section{Geometry of Concentric Black Rings}
\label{sec:ringgeometry}
Any supersymmetric solution in the asymptotically
flat $\bR^{1,4}$ is known to  be of the form \cite{allsolution}
\begin{equation}
ds^2=-f^2 (dt+\omega)^2 + f^{-1}ds^2(\bR^4)
\end{equation} where $f$ and $\omega$ is a function and a one-form
on $\bR^4$, respectively.
We parametrize the base $\bR^4$ 
in the Gibbons-Hawking coordinate system
\begin{equation}
ds^2(\bR^4)=H[dr^2+r^2(d\theta^2+\sin^2\theta d\chi^2)]+
H^{-1}(2d\psi+\cos\theta d\chi)^2
\end{equation}
where $(r,\theta,\phi)$ parametrize a flat $\bR^3$, the periodicity of $\psi$
is $2\pi$ and $H=1/r$.  Our notation mostly follows the one in Ref.~\citen{Concentric},
with the change $\psi_{\text{there}}=2\psi_{\text{here}}$. 
The quantities 
$f$, $\omega$ and the gauge field $F=dA$ are determined by
three functions $K$, $L$ and $M$ on the flat $\bR^3$.
The relations we need are 
\begin{align}
f^{-1}&=H^{-1} K^2+L, &
\iota_{\partial_\psi}\omega&=2H^{-2}K^3+3H^{-1}KL+2M,\label{eq:KLM1}\\
F&=\frac{\sqrt3}{2}d[f(dt+\omega)]-\frac1{\sqrt3}G^+,&
\iota_{\partial_\psi}G^+&=-3 d(H^{-1}K)\label{eq:KLM2}
\end{align} where $G^+=f(d\omega+\star d\omega)/2$ 
is a self-dual two-form on $\bR^4$. 

To construct the concentric black ring solutions,
we take  $N$ points $\bx_i$, ($i=1,\ldots,N)$ 
at $r=R_i^2/4,\theta=\pi$ on $\bR^3$.  The orbit of $\bx_i$
along the coordinate $\psi$ is a ring of radius $R_i$ embedded in $\bR^4$.
We choose functions $K$, $L$ and $M$ by
\begin{eqnarray}
K = - \frac12 \sum_{i=1}^N q_i h_i, \quad
L = 1+ \frac14 \sum_{i=1}^N (Q_i - q_i^2) h_i,\quad
M = \frac34 \sum_{i=1}^N q_i(1- |\bx_i|h_i )
\end{eqnarray} where  $h_i(\bx)=1/|\bx-\bx_i|$
are harmonic functions on $\bR^3$. For the case with a single ring,
conversion to the ring coordinate used in \eqref{eq:flat} 
can be achieved via \begin{equation}
\phi_1=\psi+\chi/2,\qquad \phi_2=\psi-\chi/2
\end{equation}and \begin{equation}
\frac{R\sqrt{y^2-1}}{x-y}=2\sqrt{r}\sin\frac{\theta}2,\qquad
\frac{R\sqrt{1-x^2}}{x-y}=2\sqrt{r}\cos\frac{\theta}2.
\end{equation}

The behavior of $\omega$ and $F$ at the asymptotic infinity,
and the near-horizon metric \eqref{eq:nearhorizonmetric}
are well-known and are not repeated here. The reader is referred to the original
article Ref.~\citen{Concentric}. 
The gauge potential near the horizon can be obtained by
the combination of 
\eqref{eq:KLM1} and \eqref{eq:KLM2}. First we have 
\begin{eqnarray}
\iota_{\partial_{\psi}} F = \frac{\sqrt3}2 (- d \iota_{\partial_{\psi}})[f(dt + \omega)]
+ {\sqrt3} d(KH^{-1}) .
\end{eqnarray} which can be integrated by inspection.
Hence   the $\psi$ component of the gauge field is given by
\begin{equation}
\iota_{\partial_\psi} A = \sqrt3\left[\frac{H^{-1}KL/2 +M}
{H^{-1}K^2+L}+c\right]
\end{equation} 
for some constant $c$. By demanding $\iota_\psi A\to 0$ as $r\to\infty$, we 
obtain \begin{equation}
c= - \frac12\sum_{i=1}^N q_i.
\end{equation} 
Thus, we have 
\begin{equation}
\iota_{\partial_\psi} A
=-\frac{\sqrt{3}}4 \left(\frac{Q_i-q_i^2}{q_i}+2\sum_{i=1}^N q_i\right).
\end{equation} near the $i$-th horizon.
The $\chi$  component of the gauge field is fixed by
the magnetic dipole through the horizon.

\end{document}